\documentstyle[12pt]{article}

\def\ai{\'{\i}}
\def\itt{\int_{\tau_1} ^{\tau_2}}
\def\oq{\overline Q}
\def\pb{\pi_\beta}

\def\po{\pi_\Omega}
\def\om{\Omega}
\def\e3o{e^{3\Omega}}
\def\4o2b{e^{4\Omega+2\beta}}
\def\e2ob{e^{2(\Omega-\beta)}}

\def\-3o{e^{-3\Omega}}

\def\eob{e^{\Omega-\beta}}

\def\e6ob{e^{6(\Omega+\beta)}}

\def\e3ob{e^{3(\Omega+\beta)}}
\def\be{\begin{equation}}
\def\ee{\end{equation}}

\def\px{\pi_x}
\def\ep{\epsilon}
\def\py{\pi_y}
\def\op{\overline P}
\begin{document}
\baselineskip.33in

\centerline{\large{\bf Global phase time and path integral}}

\smallskip

\centerline{\large{\bf for the Kantowski--Sachs anisotropic universe}}

\bigskip

\centerline{  Claudio Simeone\footnote{{\bf Electronic mail:} simeone@tandar.cnea.gov.ar}}

\medskip

\centerline{\it Departamento de F\ai sica, Comisi\'on Nacional de Energ\ai a At\'omica}

\centerline{\it Av. del Libertador 8250, 1429 Buenos Aires, Argentina}

\centerline{\it and}

\centerline{\it Departamento de F\ai sica, Facultad de Ciencias Exactas y Naturales}

\centerline{\it Universidad de Buenos Aires,  Ciudad Universitaria, Pabell\'on I}

\centerline{\it 1428, Buenos Aires, Argentina.}

\vskip1cm

\noindent ABSTRACT

\bigskip
 
 The action functional of the anisotropic Kantowski--Sachs cosmological model is turned into that of an ordinary gauge system. Then a global phase  time is identified for the model by imposing canonical gauge conditions,  and the quantum transition amplitude is obtained by means of the usual path integral procedure of Fadeev and Popov.  
  
\vskip1cm

{\it KEY WORDS}:\ Minisuperspace; path integral; global phase time.

\vskip1cm

{\it PACS numbers}:\  04.60.Kz\ \ \ 04.60.Gw\ \ \  98.80.Hw

\newpage

\noindent {\bf 1. INTRODUCTION}

\bigskip

 The Hamiltonian formalism for  gauge systems includes constraints $C_m$ which are linear and homogeneous in the momenta, plus a non vanishing Hamiltonian $H_0$ which is the total energy. For example, in  electromagnetism   the canonical momenta are the four quantities $F^{\mu 0}$; for $\mu=1,2,3$ we have the three components of the electric field, and for $\mu=0$ we have the linear constraint  $F^{0 0}=0$ [1].  This is not the case for  parametrized systems as the gravitational field:  the Hamiltonian  vanishes on the physical trajectories of the system, that is, we have a constraint $C={\cal H}\approx 0$ which for minisuperspace models has the form
$${\cal H}=G^{ij}p_ip_j+V(q)\approx 0,$$
where $G^{ij}$ is the reduced version of the DeWitt supermetric [2].  This reflects the fact that the evolution of the gravitational field  in General Relativity is given in terms of a parameter $\tau$ which does not have physical significance. This yields a fundamental difference between the  quantization of the gravitation and ordinary quantum mechanics, because the existence of a unitary quantum theory is  related to the possibility of defining the time as an absolute parameter. The identification of  time  can therefore be considered as the first step before quantization.
 
When quantizing constrained systems we must impose gauge conditions which must can be reached from any path in phase space by means of gauge transformations leaving the action unchanged; hence the symmetries of the action must be examined. 
Under a gauge transformation defined by the parameters $\epsilon^m$ the action of a system with constraints $C_m$ changes by [3]
$$\delta_\epsilon S=\left[ \epsilon^m (\tau )\left( p_i{\partial C_m\over\partial p_i}-C_m\right) \right]_{\tau_1} ^{\tau_2}. $$
Then for an ordinary gauge system we have $\delta_\ep S=0$, and gauge conditions of the form $\chi(q,p,\tau)=0$ ({\it canonical gauges}) are admissible. But in the case of a minisuperspace we  have a constraint that is quadratic in the momenta, so that it would be $\delta_\ep S\neq 0$ unless $\epsilon (\tau_1 )=\epsilon(\tau_2 )=0$; then  gauge conditions involving derivatives of Lagrange multipliers as, for example $\chi\equiv dN/d\tau=0,$ should be used [4,5]. This has the consequence that the usual  path integral procedure for quantizing gauge systems could not be applied. 

In a recent article [6] we studied this  problem in the case of empty Friedmann-Robertson-Walker cosmologies; we were able to find a solution which allowed us to obtain the transition amplitude by means of the usual path integral procedure of Fadeev and Popov for ordinary gauge systems. However, our procedure was rather cumbersome: We  defined a canonical transformation which matched  the cosmological models with the ideal clock [6,7]. Then we showed that the ideal clock could be turned into an ordinary gauge system by improving its action functional with gauge invariance at the end points [8]; and finally we imposed canonical gauges to find the transition amplitude for the ideal clock by means of the Fadeev-Popov method, and studied its relation with the amplitude for the minisuperspaces. An important point that we learned was that our procedure  worked only if an {\it intrinsic} time [9] (that is, a function of  the coordinates only) could be defined for the system.

In the present work we  generalize our analysis to a model with true degrees of freedom; we deparametrize the anisotropic Kantowski--Sachs [10,11] universe and show that the transition amplitude for it can be obtained by means of the usual path integral procedure for gauge systems. Here  we proceed in a more straightforward way than in Ref. 6: we solve the Hamilton-Jacobi equation for the system to obtain the generator function of a canonical transformation which turns the minisuperspace into an ordinary gauge system [12]. Then we identify an intrinsic time  by imposing a simple $\tau-$dependent canonical gauge [13], and we obtain the transition amplitude  in the form of a path integral that after the gauge choice makes apparent the  separation between true degrees of freedom and time.
We also show that with our method   an {\it extrinsic} time (i.e. a function of the coordinates and also of the momenta) can be identified for the Kantowski--Sachs universe.

\vskip1cm

\noindent {\bf 2. GAUGE-INVARIANT ACTION}

\bigskip

 Consider  the $\tau-$independent Hamilton--Jacobi equation
\be 
 G^{ij}{\partial W\over\partial q^i} {\partial W\over\partial q^j}+V(q)=E\ee
which results by substituting $p_i=\partial W/\partial q^i$ in the Hamiltonian.  A complete solution $W(q^i,\alpha_\mu ,E)$  obtained by matching the integration constants
 $(\alpha_\mu ,E)$ to  the new momenta $(\op_\mu ,\op_0)$  generates a canonical transformation [12]
\be 
p_i={\partial W\over\partial q^i},\ \ \ \ \ \ \oq^i={\partial W\over\partial \op_i},\ \ \ \ \ \  \overline K=N\op_0=N{\cal H}\ee
which identifies the constraint ${\cal H}$ with the new momentum $\op_0$. The variables $(\oq^\mu,\op_\mu)$  are conserved observables because $[\oq^\mu,{\cal H}]=[\op_\mu,{\cal H}]=0$, so that they would not be appropriate to characterize the dynamical evolution. A second transformation generated by the function
\be 
F=P_0\oq^0+f(\oq^\mu ,P_\mu ,\tau) \ee
gives
\be\op_0=P_0\ \ \ \ \ \ \  \op_\mu={\partial f\over \partial\oq^\mu}\ \ \ \ \ \ \ \
 Q^0=\oq^0\ \ \ \ \ \ \  Q^\mu={\partial f\over\partial P_\mu}\ee 
and a new non vanishing Hamiltonian
$
K=NP_0+\partial f/\partial\tau,$
so that $(Q^\mu,P_\mu)$ are non conserved observables because $[Q^\mu,{\cal H}]=[P_\mu,{\cal H}]=0$ but $[Q^\mu, K]\neq 0$ and $[P_\mu,K]\neq 0$;  we have,  instead, that $[Q^0,{\cal H}]=[Q^0,P_0]=1$, and then $Q^0$ can be used to fix the gauge [12].
The  transformation $(q^i,p_i)\  \to\ (Q^i,P_i)$ leads to the action
\be
{\cal S}[Q^i,P_i,N]=\itt \left( P_i{dQ^i\over d\tau}-NP_0-{\partial f\over\partial\tau}\right) d\tau\ee
which  contains a linear and homogeneous constraint $P_0\approx 0$ and a non zero Hamiltonian $\partial f/\partial\tau$ and is then that of an ordinary gauge system. In terms of the original  variables the gauge invariant action $\cal S$ reads 
\be
{\cal S}[q^i,p_i,N]=\itt\left( p_i{dq^i\over d\tau }-N{\cal H}\right) d\tau + B(\tau_2) -B(\tau_1)\ee
where the end point terms $B$ have the form [12]
\be B= \oq^i\op_i -W+Q^\mu P_\mu-f.\ee
As $\cal S$ and $S$ differ only in surface terms, they then yield the same dynamics.
 
We can now use the action (5) to compute the transition amplitude $<Q_2^\mu,\tau_2|Q_1^\mu,\tau_1>$ ($Q^0$ is a spurious degree of freedom for the gauge system) 
 by means of a path integral of the form
$$\int DQ^0\, DP_0\, DQ^\mu\, DP_\mu\, DN\, \delta(\chi )\,\vert [\chi,P_0]\vert\,  \exp\left[ i \itt \left( P_i{dQ^i\over d\tau}-NP_0-{\partial f\over\partial\tau}\right) d\tau\right]$$
where $\vert [\chi, P_0]\vert$ is the Fadeev-Popov determinant and canonical gauges are admissible. We want to obtain the amplitude $<q_2^i|q_1^i>$, so that we should show that both amplitudes are equivalent. This is fulfilled if the paths are weighted in the same way by ${\cal S}$ and $S$ and if $Q^\mu$ and $\tau$ define a point in the original configuration space, that is, if  a state $|Q^\mu,\tau>$ is equivalent to $|q^i>$. This is true only if there exists a gauge such that $\tau=\tau(q^i)$, and such that the surface terms (7) vanish; this dictates the choice of the function $f$ in equation (3). The existence of  a gauge condition yielding  $\tau=\tau(q^i)$ is closely related to the existence of an intrinsic time [6,12,13]; in the case of the Kantowski--Sachs cosmological model we shall see that a  gauge  of the form $\chi\equiv Q^0+T(\tau)=0$  can be used to define an extrinsic time, while an intrinsic time can be defined  by means of a gauge condition like $\chi\equiv PQ^0 + T(\tau)=0.$

\vskip1cm

\noindent {\bf 3. THE KANTOWSKI--SACHS COSMOLOGY}

\bigskip

 While isotropic Friedmann--Robertson--Walker cosmologies can be thought to be  a good description for the present universe, more general models should be considered when studying the early universe. Possible anisotropic cosmologies are comprised by the Bianchi models [14,10] and the Kantowski--Sachs model [10]. The spacetime metric of the last one can be written in the form
\be ds^2=N^2d\tau^2-e^{2\om(\tau)}\left(e^{2\beta(\tau)} d\psi^2+e^{-\beta(\tau)}(d\theta^2+\sin^2\theta d\varphi^2)\right),\ee
where $e^{2\om}$ is the spatial scale factor, and $\beta$ determines the degree of anisotropy.
In the absence of  matter the action functional reads
\be S[\om,\beta,\po,\pb,N]=\itt \left(\pb{d\beta\over d\tau}+\po{d\om\over d\tau}-N{\cal H}\right)d\tau\ee
where ${\cal H}=\-3o H$ is the Hamiltonian constraint, and
 \be H= -\po^2+\pb^2-\4o2b \approx 0.\ee
The scaled potential $v(\om,\beta)=-\4o2b$ has a definite sign, so that an intrinsic time can be identified among the  canonical variables, and the procedure of the preceding section can be applied. The Hamiltonian is not separable in terms of the original variables; then we define
$$\e3ob\equiv 3x,\ \ \ \ \ \ \ \ \eob\equiv 4y,$$
so that $v(\om,\beta)=-12xy$ and $-\po^2+\pb^2= -12xy\px\py.$ Hence we can write
$$H= -12xy(\px\py+1)\approx 0.$$
Because $xy$ is positive definite, we can define the equivalent constraint
\be H'\equiv -(\px\py+1)\approx 0.\ee
The $\tau -$ independent Hamilton-Jacobi equation associated to this constraint is 
$$-{\partial W\over\partial x}{\partial W\over\partial y} - 1=E',$$ 
and matching the integration constants $\alpha,E'$ to the new momenta $\op,\op_0$ it has the solution
\be
W(x,y,\op_0,\op)=-\op x+y\left({  1+ \op_0\over \op}\right);\ee
then
$$\px={\partial W\over\partial x}=-\op,\ \ \ \ \ \ \ \ \ \ \ \ \py={\partial W\over\partial y}={1+\op_0 \over \op},$$
$$\oq^0={\partial W\over\partial\op_0}={y\over\op},\ \ \ \ \ \ \ \ \ \ \oq= {\partial W\over\partial\op}=-x-y\left({1 +\op_0\over\op^2}\right) .$$
To go from the conserved observables $(\oq,\op)$ to  $(Q,P)$ we define
\be
F=\oq^0 P_0 + \oq P+{T(\tau)\over P}\ee
with $T(\tau)$ a monotonic function. The canonical variables of the gauge system  are therefore given by
$$P_0=-\px\py-1,\ \ \ \ \ \ \ \ \ \ \ \ \ \ P=-\px,$$
\be
Q^0={y\over P},\ \ \ \ \ \ \ \ \ \ Q=-x-\left({y(1+ P_0)+ T(\tau)\over P^2}\right) \ee
($P=-\px$ cannot be zero on the constraint surface). 
According with equation (13) the true Hamiltonian  of  the gauge system described by $(Q^i,P_i)$ is 
$h\equiv (1/ P) (dT/d\tau).$ 
Hence the gauge invariant action ${\cal S}$  can be written 
\be {\cal S}[Q^i,P_i,N]=\itt \left( P{dQ\over d\tau} +P_0{dQ^0\over d\tau}-NP_0 - {1\over P} {dT\over d\tau}\right)d\tau,\ee
or in terms of the original variables
\be {\cal S}[\om,\beta,\po,\pb,N]=\itt \left(\pb{d\beta\over d\tau}+\po{d\om\over d\tau}-N{\cal H}\right)d\tau + B(\tau_2) - B(\tau_1),
\ee
 where
\begin{eqnarray} B  & = & {1\over \po+\pb}\left[4 \e3ob\left({1\over 4}\eob+ T(\tau)\right)+{1\over 2}(-\po^2+\pb^2-\4o2b)\right]\nonumber\\
& = & -\left[ 2 \left(Q^0 + {T(\tau)\over P}\right)+Q^0 P_0 \right].
\end{eqnarray}
Under a gauge transformation generated by ${\cal H}$ we have $\delta_\epsilon B= -\delta_\epsilon S$, and hence $\delta_\epsilon {\cal S}=0.$ On the constraint surface $H'=P_0=0$ this terms clearly vanish in the gauge
\be \chi\equiv Q^0 P+ T(\tau)=0\ee
which is equivalent to
$T=-(1/ 4)\eob ,$   and then defines $\tau=\tau(\om,\beta).$  An intrinsic time $t$ can be defined   by writing $t=\eta T$, with $\eta=\pm 1$, and apropriately choosing $\eta$. A global phase time $t$ must verify $[t,{\cal H}]>0$ [15]; because ${\cal H}={\cal F}(\om,\beta) H'$ with ${\cal F}>0$, then if $t$ is a global phase time we also have $[t,H']>0$. For $t$ we have 
$$[t,H']=[-\eta Q^0 P,P_0]=-\eta P,$$
and because $P=-\px$ then we must choose $\eta= 1$ if $\px>0$ and  $\eta=-1$ if $\px<0$; as $\px=(1/2)(\po+\pb)e^{-3(\om+\beta)}$ and on the constraint surface it is $|\pb|>|\po|$, we have $sign(\px)=sign(\po+\pb)=sign(\pb)$; therefore the time is 
$$t(\om,\beta)= -T={1\over 4}\eob\ \ \ \ \ \ \ \mbox{if}\ \ \ \ \ \ \ \pb<0,$$
\be t(\om,\beta)=T=-{1\over 4}\eob\ \ \ \ \ \ \ \mbox{if}\ \ \ \ \ \ \ \pb>0.\ee
Note that $\pb$ cannot change from a negative value to a positive one on the constraint surface, so that the time is well defined for the whole evolution of the system. 

It is easy to verify that an extrinsic time can be defined by imposing a canonical gauge  of the form
$ \chi\equiv Q^0+T(\tau)=0.$ If we make $t=-T$ we  obtain
$$ t(\om,\beta,\po,\pb)  =  Q^0
 = - {e^{4\om +2\beta}\over 2(\po+\pb)}$$
 with $[t,{\cal H}]>0.$  
Using the constraint equation (10) we can write
\be t(\po,\pb)={1\over 2}(\po-\pb).\ee
We  see that a gauge condition involving one of the new momenta defines a time in terms of only the original coordinates, while a gauge involving only one of the new coordinates gives an extrinsic time which can be written in terms of only the original momenta.

Because the path integral in the variables $(Q^i,P_i)$ is gauge invariant, we can compute it in any canonical gauge. With the gauge choice (18),   on the constraint surface $P_0=0$, and after integrating on $N$, $P_0$ and $Q^0$, the transition amplitude is given by
\be <\beta_2,\om_2\vert\beta_1,\om_1 >=\int DQDP\exp\left[i\int_{T_1} ^{T_2} 
\left(PdQ-{1\over P}dT\right) \right],\ee
where the end points are given by $T_1=(1/4)e^{\om_1-\beta_1}$ and $T_2=(1/4)e^{\om_2-\beta_2}$; because on the constraint surface and in gauge (18) the true degree of freedom reduces to $Q=x=(1/3)e^{3(\om+\beta)}$, then the paths in phase space go from $Q_1=(1/3)e^{3(\om_1+\beta_1)}$ to $Q_2=(1/3)e^{3(\om_2+\beta_2)}$.
 After the gauge fixation we have obtained  the path integral  for a system with one physical degree of freedom and  with a  true Hamiltonian. The result shows the separation between true degrees of freedom and time yielding  after a simple canonical gauge choice. 

\vskip1cm

\noindent {\bf 4. CONCLUSIONS}

\bigskip

 In the theory of gravitation the Hamiltonian not only generates the dynamical evolution, but it also acts as a generator of gauge transformations which connect any pair of succesive points on each classical trajectory of the system. While the dynamics is given by a spacelike hypersurface evolving in spacetime, including arbitrary local deformations which yield a multiplicity of times, the same motion can be generated by gauge transformations [16]. It is therefore natural to think that the gauge fixing procedure can be a way to identify a global time. 

However,  as the action of  gravitation is not gauge invariant at the boundaries, this feature  could not be  used, in principle, to obtain a direct procedure for deparametrizing minisuperspaces: while ordinary gauge systems admit canonical gauges $\chi(q^i,p_i,\tau)=0$,  only derivative gauges would be admissible for cosmological models. 

Here we have shown that if we can separate the Hamilton--Jacobi equation for the model under consideration, this problem can be solved by improving the action functional with gauge invariance at the boundaries, so that canonical gauges are therefore admissible. We then have a procedure for deparametrizing the system     and, simultaneously, to obtain the quantum transition amplitude in a very simple form which clearly shows the separation between true degrees of freedom and time. We have illustrated our method with the Kantowski--Sachs universe because its anisotropy makes it  physically more interesting than usual isotropic cosmologies, and because the time is not trivially identified as a function of the scale factor. This is something to be noted, as it can sometimes be found that the isolation of the coordinate $\om$ as time parameter is made as the previous step before quantization.  This is not possible in general:  in fact, we can  see that no function $\Theta (\om)$ can be a global phase time for the Kantowski--Sachs universe:
$$[ \Theta (\om),{\cal H}]= -2{\partial \Theta (\om)\over\partial\om}e^{-3\om}\po,$$
and for $\pb=\pm e^{2\beta +\om}$ we have  $\po=0$, so that $[ \Theta (\om),{\cal H}]$ vanishes.   

A point  should be stressed, and its is  that our procedure  will work as it stands only if an intrinsic time exists, which requires a Hamiltonian constraint whose potential has a definite sign, as it is the case of the Kantowski--Sachs universe.  Other anisotropic separable models not posessing this property (as the Taub universe) would require  a further analysis.

\vskip1cm

\noindent {\bf ACKNOWLEDGMENTS}

\bigskip

I wish to thank H. De Cicco and R. Ferraro for reading  the manuscript and making helpful comments.

\newpage

\noindent {\bf REFERENCES}

\bigskip

\noindent 1. P. A. M. Dirac, {\it Lectures on Quantum Mechanics,} Belfer Graduate School of Science, Yeshiva University, New York (1964).

\noindent 2. J. J. Halliwell, in {\it Introductory Lectures on Quantum Cosmology}, Proceedings of the Jerusalem Winter School on Quantum Cosmology and Baby Universes, edited by T. Piran, World Scientific, Singapore (1990).

\noindent 3. M. Henneaux  and C. Teitelboim, {\it Quantization of Gauge Systems}, Princeton University Press, New Jersey (1992).

\noindent 4. C. Teitelboim, Phys. Rev. D {\bf 25}, 3159 (1982).
 
\noindent 5. J. J. Halliwell, Phys. Rev. D {\bf 38}, 2468 (1988).

\noindent 6. H. De Cicco and C. Simeone, Gen. Rel. Grav. {\bf 31}, 1225 (1999).

\noindent 7.  S.  C.   Beluardi and R.  Ferraro, Phys. Rev. D {\bf 52}, 1963 (1995).

\noindent 8. M. Henneaux, C. Teitelboim  and J. D. Vergara, Nucl. Phys. B {\bf 387}, 391 (1992).

\noindent 9. K. V. Kucha\v r, in {\it Proceedings of the 4th Canadian Conference on General Relativity and Relativistic Astrophysics}, edited by G. Kunstatter, D. Vincent and J. Williams, World Scientific, Singapore  (1992).

\noindent 10. M. P. Ryan and L. C. Shepley, {\it Homogeneous Relativistic Cosmologies},
Princeton Series in Physics, Princeton University Press, New Jersey (1975).

\noindent 11. A. Higuchi and R. M. Wald, Phys. Rev, D {\bf 51}, 544 (1995)

\noindent 12. R. Ferraro  and C. Simeone, J. Math. Phys, {\bf 38}, 599 (1997).

\noindent 13. C. Simeone, J. Math. Phys. {\bf 40}, 4527 (1999).

\noindent 14. L. D. Landau and E. M. Lifshitz, {\it The Classical Theory of Fields}, Pergamon Press, Oxford (1975).

\noindent 15. P. H\'aj\ai cek, Phys. Rev. D {\bf 34}, 1040 (1986).

\noindent 16. A. O. Barvinsky, Phys. Rep. {\bf 230}, 237 (1993).

\end{document}